\documentclass[preprint,aps,newabstract]{revtex4}

\usepackage{graphicx}
\usepackage{rotating}
\usepackage{axodraw}
\usepackage{latexsym}

\newcommand{\lsim}{\mathrel{\mathop{\kern 0pt \rlap
  {\raise.2ex\hbox{$<$}}}
  \lower.9ex\hbox{\kern-.190em $\sim$}}}
\newcommand{\gsim}{\mathrel{\mathop{\kern 0pt \rlap
  {\raise.2ex\hbox{$>$}}}
  \lower.9ex\hbox{\kern-.190em $\sim$}}}
\newcommand{\beq}     {\begin{equation}}
\newcommand{\eeq}     {\end{equation}}
\newcommand{\bea}     {\begin{eqnarray}}
\newcommand{\eea}     {\end{eqnarray}}

\newcommand{\sg}      {\sigma}

\newcommand{\no}      {\nonumber}

\newcommand{\nc}{\newcommand}
\nc{\postscript}[2]
{\setlength{\epsfxsize}{#2\hsize}\centerline{\epsfbox{#1}}}
\nc{\bg}{B. Grzadkowski}
\nc{\non}{\nonumber}
\nc{\hc}{\hbox {h.c.}} \nc{\re}{\hbox {Re}} 
\nc{\mev}{\hbox {MeV}} \nc{\gev}{\;\hbox {GeV}} \nc{\tev}{\;\hbox {TeV}}
\def\lsim{\mathrel{\raise.3ex\hbox{$<$\kern-.75em\lower1ex\hbox{$\sim$}}}}
\def\gsim{\mathrel{\raise.3ex\hbox{$>$\kern-.75em\lower1ex\hbox{$\sim$}}}}

\nc{\Lsp}{\;\;\;\;\;\;\;\;\;\;}  \nc{\LLLsp}{\lspace \lspace}
\nc{\lsp}{\;\;\;\;\;\;}
\nc{\spac}{\;\;\;}
\nc{\noi}{\noindent}
\nc{\baa}{\begin{array}}      \nc{\eaa}{\end{array}}
\nc{\bit}{\begin{itemize}}    \nc{\eit}{\end{itemize}}
\nc{\ben}{\begin{enumerate}}  \nc{\een}{\end{enumerate}}
\nc{\bce}{\begin{center}}     \nc{\ece}{\end{center}}

\def\Hhat{\widehat H}
\def\vo{v_0}
\def\ho{h_0}
\def\mho{m_{\ho}}
\def\phio{\phi_0}
\def\mphio{m_{\phio}}

\def\anti{\overline}

\def\gam{\gamma}

\def\mphi{m_\phi}

\def\lwh{\widehat\Lambda_W}

\def\lphi{\Lambda_\phi}

\def\mphi{m_\phi}

\def\hbar{\overline h}
\def\grav{h_{\mu\nu}^{(n)}}
\def\lam{\lambda}
\def\Lam{\Lambda}
\def\mpl{M_{\rm Pl}}
\def\ifmath#1{\relax\ifmmode #1\else $#1$\fi}
\def\half{\ifmath{{\textstyle{1 \over 2}}}}

\def\call{{\cal L}}

\def\sig{\sigma}
\def\eps{\epsilon}

\begin{document}

\renewcommand{\thepage}{-- \arabic{page} --}
\def\mib#1{\mbox{\boldmath $#1$}}
\def\bra#1{\langle #1 |}      \def\ket#1{|#1\rangle}
\def\vev#1{\langle #1\rangle} \def\dps{\displaystyle}
\nc{\tb}{\stackrel{{\scriptscriptstyle (-)}}{t}}
\nc{\bb}{\stackrel{{\scriptscriptstyle (-)}}{b}}
\nc{\fb}{\stackrel{{\scriptscriptstyle (-)}}{f}}
\nc{\pp}{\gamma \gamma}
\nc{\pptt}{\pp \to \ttbar}
\nc{\barh}{\overline{h}}
   \def\thebibliography#1{\centerline{REFERENCES}
     \list{[\arabic{enumi}]}{\settowidth\labelwidth{[#1]}\leftmargin
     \labelwidth\advance\leftmargin\labelsep\usecounter{enumi}}
     \def\newblock{\hskip .11em plus .33em minus -.07em}\sloppy
     \clubpenalty4000\widowpenalty4000\sfcode`\.=1000\relax}\let
     \endthebibliography=\endlist
   \def\sec#1{\addtocounter{section}{1}\section*{\hspace*{-0.72cm}
     \normalsize\bf\arabic{section}.$\;$#1}\vspace*{-0.3cm}}
\preprint{ \hbox{\bf hep-ph/0301002}}
\preprint{ \hbox{NSC-NCTS-030101}} \preprint{ \hbox{KIAS-P02085}}

\vskip 1cm

\title{A probe of the Radion-Higgs mixing \\
in the Randall-Sundrum model
at $e^+ e^-$ colliders}

\author{
Kingman Cheung}
\affiliation{National Center for Theoretical Sciences, National
Tsing Hua University, Hsinchu, Taiwan}
\author{
C.~S. Kim}
\affiliation{Department of Physics and
IPAP, Yonsei University, Seoul 120-749, Korea}
\author{
Jeonghyeon Song}
\affiliation{School of Physics,
Korea Institute for Advanced Study\\
207-43 Cheongryangri-dong,
Dongdaemun-gu, Seoul 130-012, Korea}

\begin{abstract}
In the Randall-Sundrum model,
the radion-Higgs mixing is weakly suppressed
by the effective electroweak scale.
A novel feature of the existence of gravity-scalar mixing
would be a sizable three-point vertex, $\grav$-$h$-$\phi$.
We study this vertex in the process
$e^+ e^- \to\grav \to h \phi$, which is allowed only with
a non-zero radion-Higgs mixing.
It is shown that
the angular distribution is a unique characteristic of
the exchange of massive spin-2 gravitons,
and the total cross section at the future $e^+ e^-$ colliders
is big enough
to cover a large portion of the parameter space
where the LEP/LEP II data cannot constrain.
\end{abstract}

\maketitle

\section{Introduction}

Although the standard model (SM) has been
very successful in describing the electroweak
interactions
of the gauge bosons and fermions,
an important ingredient, the Higgs boson,
awaits
to be experimentally discovered\,\cite{Higgs}.
In the SM, the Higgs boson plays a central role
of the electroweak
symmetry breaking.
Since its mass should be lighter than
the scale $(8\pi \sqrt{2}/3 G_F)^{1/2}\sim 1$ TeV
for the preservation of unitarity
in the $W_L W_L \to W_L W_L$ process\,\cite{Duncan:1986vj},
the primary
efforts of future collider experiments are directed
toward the search for the Higgs boson.

Radiative corrections to the mass of the Higgs boson give
rise to the gauge hierarchy problem, which
has motivated a number of models for physics beyond the SM.
Recently, it was known that the enormous hierarchy between
the electroweak and Planck scale can be explained,
without resort to any new symmetry, by introducing
extra dimensional space\,\cite{ADD,RS}.
In particular, a scenario proposed by Randall and Sundrum (RS)\,\cite{RS}
has drawn a lot of interests, in which
an additional spatial dimension of a $S^1/Z_2$ orbifold is introduced
with two 3-branes at the fixed
points.  A geometrical suppression factor, called the warp factor,
can naturally explain the gauge hierarchy
with moderate values of the model parameters.
The original RS model has a four-dimensional massless scalar field, the modulus or
radion, about the background geometry.
In order to avoid unconventional
cosmological equations,
a stabilization mechanism is required\,\cite{GW,Csaki-cosmology},
through
which the brane configuration is stabilized and the radion attains a mass.
Moreover, the radion
is likely to be lighter than the Kaluza-Klein states of any bulk fields.
Therefore, the radion will probably be the first sign of the warp-geometry.

Various phenomenological aspects of the radion
have been studied in the literature, including
its decay modes\,\cite{Ko,Wells},
its effects
on the oblique parameters
of the electroweak precision observations\,\cite{Csaki-EW},
and other effects on phenomenological signatures
at present and future colliders\,\cite{collider}.
Above all, a possible mixing between the radion and Higgs boson
may cause some deviations in phenomenological observables of the Higgs boson
from the SM ones
even in the minimal scenario in which all the SM fields
are confined on the TeV brane.
The mixing can come from the simplest example of gravity-scalar mixing,
$\xi
R(g_{\rm vis}) \widehat{H}^\dagger \widehat{H}$,
where $R(g_{\rm vis})$ is the Ricci scalar of the induced metric
$g_{\rm vis}^{\mu\nu}$
and $\widehat{H}$ is the Higgs field in the five-dimensional context.
Some studies of the effects of the radion-Higgs mixing have also been
performed,
$e.g.$, the production and decay of scalar particles at
the LHC\,\cite{Datta-HR-LHC}, the unitarity bounds\,\cite{Han-unitarity},
and the total and partial decay widths
of the Higgs boson\,\cite{Hewett:2002nk}.

Another interesting way to probe the radion-Higgs mixing
is to search for new couplings emerging as the mixing turns on:
Good examples are the triple vertices linear in the Higgs field.
In Ref.~\cite{Gunion},
it was shown that
in the effective potential for the SM Higgs boson interacting
with the KK gravitons and the radion, but without the radion-Higgs mixing,
the SM minimum of
$\left.\partial V_h /\partial h_0 \right|_{\langle h_0 \rangle =0} = 0$
is a unique one, where $h_0$ is the SM Higgs boson without the mixing.
Therefore, the vertices linear in the Higgs field $h_0$ are absent
when the radion-Higgs mixing vanishes.
As the radion-Higgs mixing turns on,
non-zero vertices of $h$-$\phi$-$h_{\mu\nu}^{(n)}$ and $h$-$\phi$-$\phi$
are generated, where $h$ ($\phi$) is physical Higgs (radion) state
and $\grav$ is the KK graviton.
The vertex of $h$-$\phi$-$\phi$ has been examined in the decay of
$h \to \phi\phi$, which can be sizable in some
parameter space\,\cite{Gunion}.
However, investigating a vertex in a specific decay mode
depends crucially on the mass spectrum of the Higgs boson and
the radion. The decay $h\to \phi\phi$ is only possible when $m_h>2 \mphi$.
Moreover,
the branching ratio of $h\to\phi\phi$ for a
light radion (e.g., for $m_\phi= 30$ GeV and $m_h=120$ GeV)
is below $10^{-3}$, which may be too small for detection.

Instead, here we focus on the $h$-$\phi$-$h_{\mu\nu}^{(n)}$
vertex by studying the associated production of
the radion with the Higgs boson at the future $e^+ e^-$ colliders.
As shall be shown below,
this high-energy  process has several advantages:
(i) the observation of this rare process would exclusively probe
the radion-Higgs mixing,
(ii) the angular distribution could reveal the exchange of massive KK
gravitons,
(iii) the coupling strength of $h$-$\phi$-$h_{\mu\nu}^{(n)}$
is much larger than the other KK-graviton-involved coupling
$\phi$-$\phi$-$h_{\mu\nu}^{(n)}$, which also vanishes as $\xi\to 0$, and
(iv) the SM background of $e^+ e^-\to b \bar{b} b \bar{b}$ is
small enough to easily detect the signal.

This paper is organized as follows.
Section \ref{review} summarizes
the RS model and the basic properties of the radion-Higgs mixing.
In Sec.~\ref{result},
the process of $e^+ e^- \to h_{\mu\nu}^{(n)}\to h \phi$
is studied in detail.
Section \ref{conclusion} deals with the summary and conclusions.

\section{Review of the Randall-Sundrum model and radion-Higgs mixing}
\label{review}

In the RS scenario, a single extra dimension is introduced
with non-factorizable geometry, which is compactified on a
$S^1/Z_2$ orbifold with size $b_0$\,\cite{RS}.
Two orbifold fixed points accommodate two three-branes,
the Planck brane at $y=0$ and
our visible brane at $y=1/2$.
If the bulk cosmological constant $\Lambda$ and the brane
cosmological constants $V_{hid,\,vis}$
satisfy the relation of
$\Lambda/m_0=-V_{\rm hid}=V_{\rm vis}=-12m_0/\eps^2$,
the following classical solution
to Einstein equations respects
the four-dimensional Poincare invariance:
\beq
ds^2=e^{-2\sigma(y)}\eta_{\mu\nu}dx^\mu dx^\nu-b_0^2dy^2,
\label{metricz}
\eeq
where
$\eta_{\mu\nu}$ is the Minkowski metric,
$\sigma(y)=m_0 b_0|y|$, and $y \in [0,1/2]$.
Here the five-dimensional Planck mass $M_5$ is denoted by
$\eps \equiv 1/M_5^3$.
The four-dimensional Planck mass on the visible brane,
defined by $M_{\rm Pl} \equiv 1/\sqrt{8\pi G_N}$,  is
\beq
{M_{\rm Pl}^2\over 2}={1-\Omega_0^2  \over \eps^2 m_0},
\label{mplkappam0}
\eeq
where $\Omega_0\equiv e^{-m_0 b_0/2}$ is known as the warp factor.
Since our brane is arranged to be at $y=1/2$,
a canonically normalized scalar field has  the mass
multiplied by the warp factor, $i.e.$, $m_{phys}=\Omega_0 m_0$.
Since the moderate value of $m_b b_0/2 \simeq 35$
can generate TeV scale physical mass,
the gauge hierarchy problem is explained.

In the original RS scenario, the compactification radius
$b_0$ is assumed to be constant:
No quantum fluctuation about the extra-dimension size
was considered\,\cite{RS}.
However, the cosmological evolution in the RS scenario
requires a fine-tuning between the densities
on the two branes, otherwise the
two branes blow apart, i.e., $b_0 \to \infty$.
The problem is that this fine-tuning leads to
unconventional cosmology\,\cite{Csaki-cosmology}.
It is shown that a stabilization mechanism
can naturally avoid this problematic fine-tuning,
inducing a massive scalar
corresponding to
the gravitational fluctuation in the distance
between two branes.

In the minimal RS model, in which
all the SM fields are confined on
the visible brane,
new phenomenological ingredients are
two kinds of gravitational fluctuations
about the RS metric:
\beq
\eta_{\mu\nu} \to \eta_{\mu\nu}+\epsilon h_{\mu\nu}(x,y)
,\lsp b_0\to b_0+b(x)\,.
\label{metric}
\eeq
The five-dimensional gravitational field is expanded
as a sum of KK modes given by
\beq
h_{\mu\nu}(x,y) = \sum_{n=0}^\infty
h_{\mu\nu}^{(n)}(x)\frac{\chi^{(n)}(y)}{\sqrt{b_0}}
\,,
\eeq
and the canonically normalized radion field, $\phi_0(x)$,
is
\beq \phio(x) \equiv
\left({12 \over \eps^2 m_0}\right)^{1/2}\Omega_b(x) \simeq
\sqrt{6}\mpl\Omega_b(x)\,,
\label{radiondef}
\eeq
where
\beq
\Omega_b(x)\equiv e^{-m_0[b_0+b(x)]/2}
\,.
\eeq
The compactification of the fifth dimension
yields
the four-dimensional interaction Lagrangian of the KK gravitons
and the radion as
\beq
{\cal L}=-\frac{\phi_0}{\Lambda_\phi} T_\mu^\mu
-\frac{1}{\lwh}  T^{\mu\nu}(x) \sum_{n=1}^\infty
h^{(n)}_{\mu\nu}(x)
\,,
\eeq
where $\Lambda_\phi$ is the vacuum expectation value (VEV)
of the radion field, $T^\mu_\mu$ is the trace of the symmetric
energy-momentum tensor $T^{\mu\nu}$,
and $\lwh = \sqrt{2} \mpl \Omega_0$.

Note that
all the known gauge and discrete symmetries of the SM
as well as the Poincare invariance on the visible brane
do not prohibit the following gravity-scalar mixing\,\cite{Wells,Gunion}:
\beq
S_\xi=\xi \int d^4 x \sqrt{g_{\rm vis}}R(g_{\rm vis})\Hhat^\dagger \Hhat\,,
\eeq
where $R(g_{\rm vis})$ is the Ricci scalar for the induced metric
on the visible brane,
$g^{\mu\nu}_{\rm vis}=\Omega_b^2(x)(\eta^{\mu\nu}+\eps h^{\mu\nu})$,
$\Hhat$ is the Higgs field before re-scaling,
i.e., $H_0=\Omega_0 \Hhat$, and
$\xi$ quantifies the size of the mixing term.
The free-field Lagrangian of the Higgs boson and radion
is given by\,\cite{Gunion}
\beq
\call_0=-\half\left\{1+6\gamma^2 \xi \right\}\phi_0\Box\phi_0
-\half\phi_0 \mphio^2\phi_0-\half h_0 (\Box+\mho^2)h_0-6\gamma \xi \phi_0\Box h_0\,,
\label{keform}
\eeq
where
\beq \gamma\equiv \vo/\lphi\,,
\label{gamdef}
\eeq
and $v_0$ is the VEV of the Higgs boson around 246 GeV.

We introduce the states $h$ and $\phi$
which diagonalize $\call_0$, defined by
\bea
\label{matrix}
\left(
\begin{array}{c}
  h_0 \\
  \phi_0 \\
\end{array}
\right)
&=&
\left(
\begin{array}{cc}
  1 & 6 \xi \gamma/Z \\
  0 & -1/Z \\
\end{array}
\right)
\left(
\begin{array}{rc}
  \cos\theta & \sin\theta \\
  -\sin\theta & \cos\theta \\
\end{array}
\right)
\left(\begin{array}{c}
  h \\
  \phi \\
\end{array}
\right) \\ \label{def-matrix}
&\equiv&
\left(
\begin{array}{cc}
  d & c \\
  b & a \\
\end{array}
\right)
\left(\begin{array}{c}
  h \\
  \phi \\
\end{array}
\right)
\,,
\eea
where
\beq
Z^2\equiv 1+6\xi\gam^2(1-6\xi)\equiv \beta-36\xi^2\gam^2\,.
\label{z2}
\eeq
The first matrix in Eq.~(\ref{matrix}) diagonalizes
the kinetic terms of the Lagrangian, which leads to the mass matrix of
\beq
\call_m = -\frac{1}{2}
\left(
\begin{array}{cc}
  h' & \phi' \\
\end{array}
\right)
\left(\begin{array}{cc}
  m_{h_0}^2 & 6\xi\gamma m_{h_0}^2 /Z \\
  6\xi\gamma m_{h_0}^2 /Z &(m_{\phi_0}^2+36\xi^2\gamma^2m_{h_0}^2)/Z^2 \\
\end{array}
\right)
\left(
\begin{array}{c}
  h' \\ \phi' \\
\end{array}
\right)
.
\eeq
This symmetric mass matrix is further diagonalized
by an orthogonal mass matrix
with the mixing angle $\theta$ given by
\beq
\tan2\theta
=12 \gam \xi Z {\mho^2\over \mphio^2-\mho^2(Z^2-36\xi^2\gam^2)}
\,.
\eeq
Note that in the RS scenario
the radion-Higgs mixing is
only suppressed by the $1/\Lam_\phi$, which
is  of order of the electroweak scale, while
in the large extra-dimensional model
the mixing is severely suppressed by the Planck scale.

The eigenvalues for the square of masses are
\beq
m_\pm^2={1\over 2 Z^2}\left\{\mphio^2+\beta \mho^2\pm\sqrt{
(\mphio^2+\beta \mho^2)^2-4Z^2\mphio^2\mho^2}\right\}
\label{emasses}
\,,
\eeq
where $m_+$ is the larger of the Higgs mass $m_h$ and the
radion mass $m_\phi$.
Since the mixing matrix in Eq. (\ref{def-matrix})
is not unitary, there is an ambiguity which particle
should be called the Higgs or radion.
In the following,
$m_{h_0}$ is set to be the Higgs mass in the limit of $\xi\to 0$.
We refer the reader to Ref.~\cite{Gunion} for the detailed
recipe to obtain $m_{h_0}$, $m_{\phi_0}$, $\theta$
(thus $a$, $b$, $c$, and $d$ in Eq.~(\ref{def-matrix}))
from the given $\gamma$, $\xi$, $m_h$ and $m_\phi$.

There are two algebraic constraints on the value of $\xi$.
One is from the requirement that the square root of the inverse function
of Eq.~(\ref{emasses}) is
positive definite, leading to
\beq
{m_+^2\over m_-^2}>1+{2\beta\over Z^2}
\left(1-{Z^2\over\beta}\right)
+{2\beta\over Z^2}\left[1-{Z^2\over \beta}\right]^{1/2}\,.
\label{rootconstraint}
\eeq
The second one comes from the condition $Z^2>0$:
\beq
\frac{1}{12}\left(1-\sqrt{1+\frac{4}{\gamma^2}}\right)
\leq \xi \leq
\frac{1}{12}\left(1+\sqrt{1+\frac{4}{\gamma^2}}\right)\,.
\label{xilim}
\eeq

All phenomenological signatures of the RS model
including the radion-Higgs mixing are specified by
five parameters
\beq
\label{parameter}
\xi,\quad \lphi,\quad \frac{m_0}{\mpl},\quad m_\phi,\quad
m_h
\,,
\eeq
which in turns determine $\lwh$ and KK graviton masses $m_G^{(n)}$ as
\bea
\label{parameter-relation}
\lwh = \frac{\lphi}{\sqrt{3}}
, \quad
m_G^{(n)} = x_n \frac{m_0}{\mpl} \frac{\lwh}{\sqrt{2}}
\,.
\eea
Here $x_n$ is the $n$-th root of the first order Bessel function.

Some comments on the parameters in Eq.~(\ref{parameter}) are in order here.
First the dimensionless coefficient of the radion-Higgs mixing,
$\xi$, approaches 1/6 in the conformal limit with $m_h \to 0$.
In general, the $\xi$ is expected to be of order one.
In most of the parameter space,
Eq.~(\ref{rootconstraint})
is crucial to bound $\xi$.
The $\Lam_\phi$  is also constrained
since it is related
by Eq.~(\ref{parameter-relation}) to
the masses and effective couplings
of KK gravitons.
The Tevatron Run I data of Drell-Yan process
and the electroweak precision data
were analyzed to constrain $m_G^{(1)}\gsim 600$ GeV,
which corresponds $\lphi\gsim 4$ TeV\,\cite{RS-onoff}.
A reasonable range of the ratio $m_0/\mpl$
is believed to be $0.01 \lsim m_0/\mpl \lsim 0.1$
since a large value of $m_0/\mpl$ would yield a large bulk curvature
damaging
the reliability of the RS solution\,\cite{Hewett-bulk-gauge}.
In what follows, we consider the case of $
\lphi=5$ TeV and $m_0/\mpl \sim 0.1$
where
the effect of radion on the oblique parameters
is small\,\cite{Csaki-EW}.
About the Higgs boson mass,
we could wonder if the radion-Higgs mixing
can weaken the $Z$-$Z$-$h$ coupling such that
the LEP/LEP II may miss the Higgs boson with a mass
below the current bound of 113 GeV.
The answer is mostly negative
due to an exact sum rule  that
the sum of the squares of $Z$-$Z$-$h$ and $Z$-$Z$-$\phi$ couplings
should be larger than the SM one\,\cite{Han-unitarity}.
Both couplings cannot be suppressed.
For $m_h=110$ GeV,
LEP/LEP II data on the Higgs search
exclude a large part of the parameter space
while for $m_h=120$ GeV
they allow most of the parameter space\,\cite{Gunion}.
In the following, we safely take the Higgs mass to be $120$ GeV.
The mass scale of the radion depends on a specific stabilization mechanism.
As the simplest mechanism by
Goldberger and Wise predicts the relation $m_{\phi_0} \sim \lwh/40$,
the radion is generically light\,\cite{GW}.
We notice that the decay mode of $h\to \phi\phi$,
exclusively allowed for $\xi \neq 0$,
can have sizable branching ratios for $m_\phi=40 \sim 60$ GeV if $m_h=120$.
In order to provide complementary information
through the process $e^+e^-\to h \phi$,
we consider the cases of $m_\phi=30,~70$ and $170$ GeV.

\begin{figure}[t!]
\begin{center}
\begin{picture}(700,50)(10,0)
\Text(15,45)[]{$h^{(n)}_{\mu\nu}$}
\Text(15,25)[]{$k_3$}
\Text(65,65)[]{$h$}
\Text(65,45)[]{$k_1$}
\Text(65,25)[]{$\phi$}
\Text(65,5)[]{$k_2$}
\DashArrowLine(40,35)(15,35){3}
\DashArrowLine(40,35)(65,55){3}
\Gluon(40,35)(15,35){2}{3}
\DashArrowLine(40,35)(65,15){3}
\Text(85,35)[l]{$i \frac{4 k_{1\,\mu}k_{2\,\nu}
}{\lwh} g_{_{G h \phi}}\equiv$}
\Text(95,35)[l]{ $\phantom{{\anti g_{n\phi h}}k_{1\,\mu}k_{2\,\nu}:}
{i}{ 4k_{1\,\mu}k_{2\,\nu} \over \lwh}
\Bigl\{3\gam\xi\left[a(\gam b+d)+bc\right]+\half cd
\Bigr\}$}
\end{picture}
\end{center}
\begin{center}
\begin{picture}(700,50)(10,0)
\Text(15,45)[]{$h^{(n)}_{\mu\nu}$}
\Text(15,25)[]{$k_3$}
\Text(65,65)[]{$\phi$}
\Text(65,45)[]{$k_1$}
\Text(65,25)[]{$\phi$}
\Text(65,5)[]{$k_2$}
\DashArrowLine(40,35)(15,35){3}
\Gluon(40,35)(15,35){2}{3}
\DashArrowLine(40,35)(65,55){3}
\DashArrowLine(40,35)(65,15){3}
\Text(85,35)[l]{$i \frac{4 k_{1\,\mu}k_{2\,\nu} }{
\lwh} g_{_{G \phi \phi}}\equiv$}
\Text(95,35)[l]{ $\phantom{{\anti g_{n\phi \phi}}k_{2\,\nu}k_{3\,\mu}:}
{i}{4 k_{1\,\mu}k_{2\,\nu}
\over\lwh} \Bigl\{3a\gam\xi\left[a\gam+2c\right]+\half c^2
\Bigr\}$}
\end{picture}
\end{center}
\caption{\it Feynman rules for the tri-linear vertices
involving $h_{\mu\nu}^n$, where
we have made use of the symmetry of $h_{\mu\nu}^n$ under
the interchange of $\mu\leftrightarrow\nu$ indices.
\label{Feynman-diagram}
}
\end{figure}
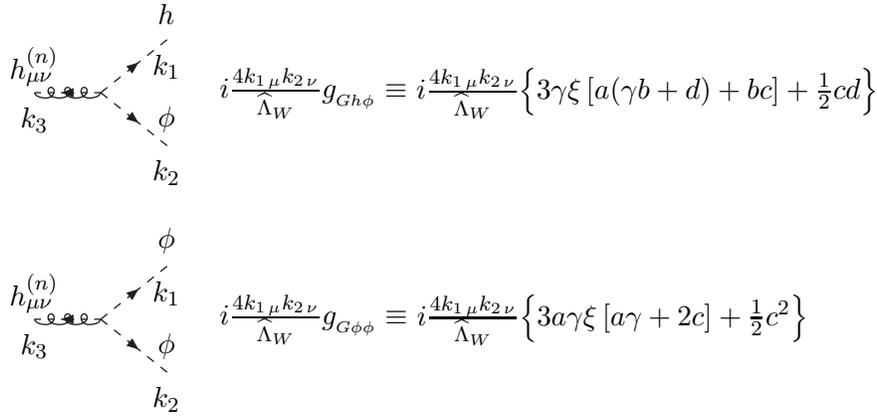

The gravity-scalar mixing $\xi\, R\, \Hhat^\dagger \Hhat$
modifies the couplings among the $h$, $\phi$ and $\grav$.
In particular,
the following four tri-linear vertices emerge, which would vanish
if the mixing goes to zero,
\beq
\label{4vertices}
h\,\mbox{-}\,\phi\,\mbox{-}\,\phi, \quad
\grav\,\mbox{-}\,h\,\mbox{-}\,\phi,\quad
\phi\,\mbox{-}\,\phi\,\mbox{-}\,\phi,\quad
\grav\,\mbox{-}\,\phi\,\mbox{-}\,\phi
\,.
\eeq
Without the radion-Higgs mixing,
the first two couplings linear in the Higgs field
are prohibited since the unique minimum of the effective potential of
the Higgs boson is the SM one, i.e.,
$\left.\partial V_h /\partial h_0 \right|_{\langle h_0 \rangle =0} = 0$.
The triple coupling $\phi$-$\phi$-$\phi$,
though allowed in a stabilization mechanism,
is suppressed by a factor of $m_{\phi}/\mpl$\,\cite{Gunion}.
The vertex of $h$-$\phi$-$\phi$ has been investigated in
the decay $h \to \phi\phi$ \cite{Gunion}.
Apart from the fact that the measurement of this decay mode
is only possible  when $m_h>2 \mphi$,
the branching ratio of $h\to\phi\phi$ for
a light radion ($m_\phi \lsim 30$ GeV and $m_h=120$ GeV)
is too small for detection.
Therefore, it is important to study the
scattering processes
to uniquely probe the other vertices in Eq.~(\ref{4vertices}),
especially those involving the KK gravitons.
In Fig.~\ref{Feynman-diagram},
we present the Feynman rules for the vertices
$\grav$-$\phi$-$\phi$ and
$\grav$-$h$-$\phi$.

\begin{figure}
\includegraphics{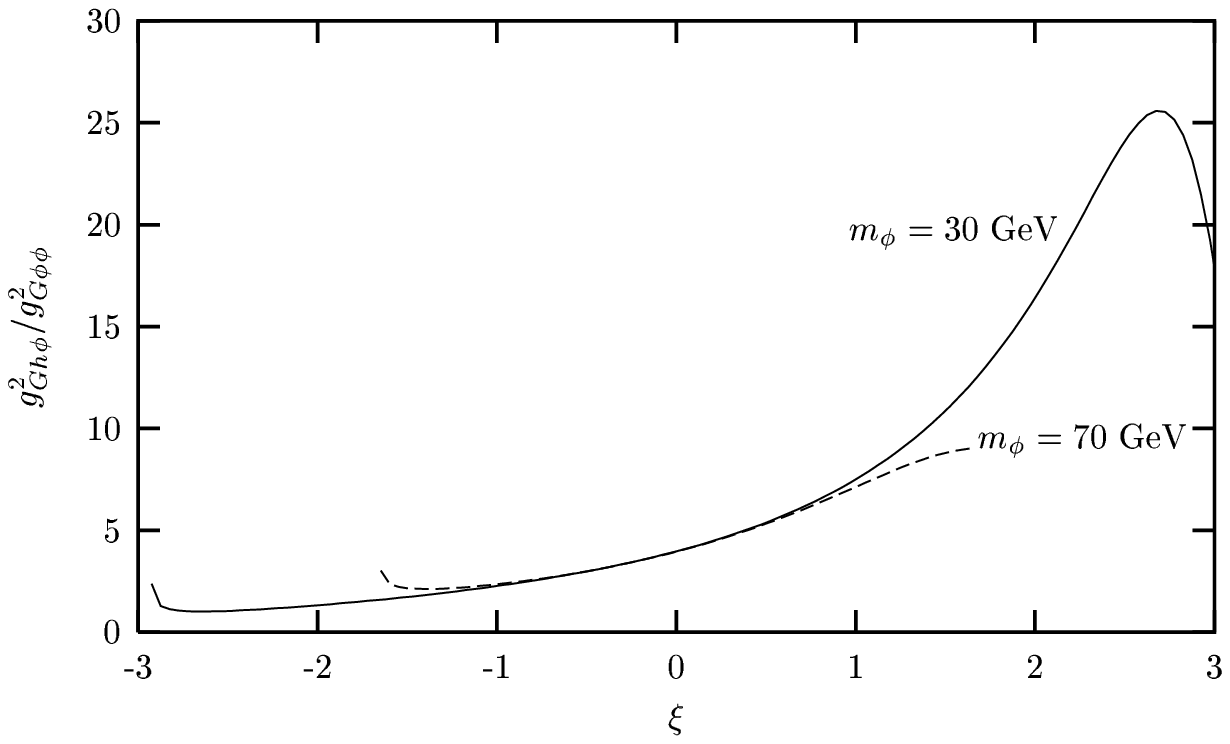}
\caption{\label{g} The ratio of $g_{_{G h \phi}}^2$
to $g_{_{G \phi \phi}}^2$
as a function of $\xi$ with $\Lam_\phi=5$ TeV and $m_0/\mpl=0.1$.}
\end{figure}

In most of the parameter space,
the coupling strength of $\grav$-$h$-$\phi$ is much larger than that of
$\grav$-$\phi$-$\phi$.
This can be easily shown by noticing that
the parameter $\gam \equiv v_0/\lphi$
 is very small with $\lphi=5$ TeV.
In the limit of $
\gam \ll 1$, we have
$
a,d \sim \mathcal{O}(1)$ but
$
b,c \sim \mathcal{O}(\gam)$,
implying
\beq
g_{_{G h \phi}} \sim \mathcal{O}(\gam),
\quad
g_{_{G \phi \phi}} \sim \mathcal{O}(\gam^2).
\eeq
Figure \ref{g} shows the ratio of $g_{_{G h \phi}}^2$ to $g_{_{G \phi \phi}}^2$
as a function of $\xi$, for $\mphi=30$ and $70$ GeV.
The $g_{_{G h \phi}}^2$ is at least a few times to more than an order of
magnitude larger than $g_{_{G \phi \phi}}^2$.
Even though
numerically $g_{_{G \phi \phi}}$ with $\Lam_\phi=5$ TeV ($\gam\simeq 0.05$)
is still not negligible,
the $\grav$-$h$-$\phi$ vertex
has largest chance to probe
the radion-Higgs mixing
through high-energy scattering processes.

\section{Numerical analysis on $e^+ e^- \to h_{\mu\nu}^{(n)} \to h \phi$}
\label{result}

For the process
\beq
e^-(p_1,\lam_{e^-}) +  e^+(p_2,{\lam}_{e^+}) \rightarrow h(k_1)+\phi(k_2)
\,,
\eeq
the helicity amplitudes $\mathcal{M}(\lam_{e^-},{\lam}_{e^+})$ are
\bea
\mathcal{M}(++)\!\!&=&\!\!\mathcal{M}(--)=0,
\\ \no
\mathcal{M}(+-)\!\!&=&\!\!\mathcal{M}(-+)=
\frac{g_{_{G h \phi}}}{2}
\sum_n \frac{1}{1 - m_{(n)}^2/s}
\frac{s}{\lwh^2} \beta \sin 2 \Theta
\,.
\eea
Here $\lam_{e^-}$ ($\lam_{e^+}$)
is the helicity of the electron (positron),
$\beta = 1+ \mu_h^4+\mu_\phi^4- 2\mu_h^2 -2 \mu_\phi^2 -2 \mu_h^2 \mu_\phi^2$,
$\mu_{h,\phi} \equiv m_{h,\phi}/\sqrt{s}$,
and $\Theta$ is the scattering angle of the Higgs boson
with respect to the electron beam.

\begin{figure}
\includegraphics{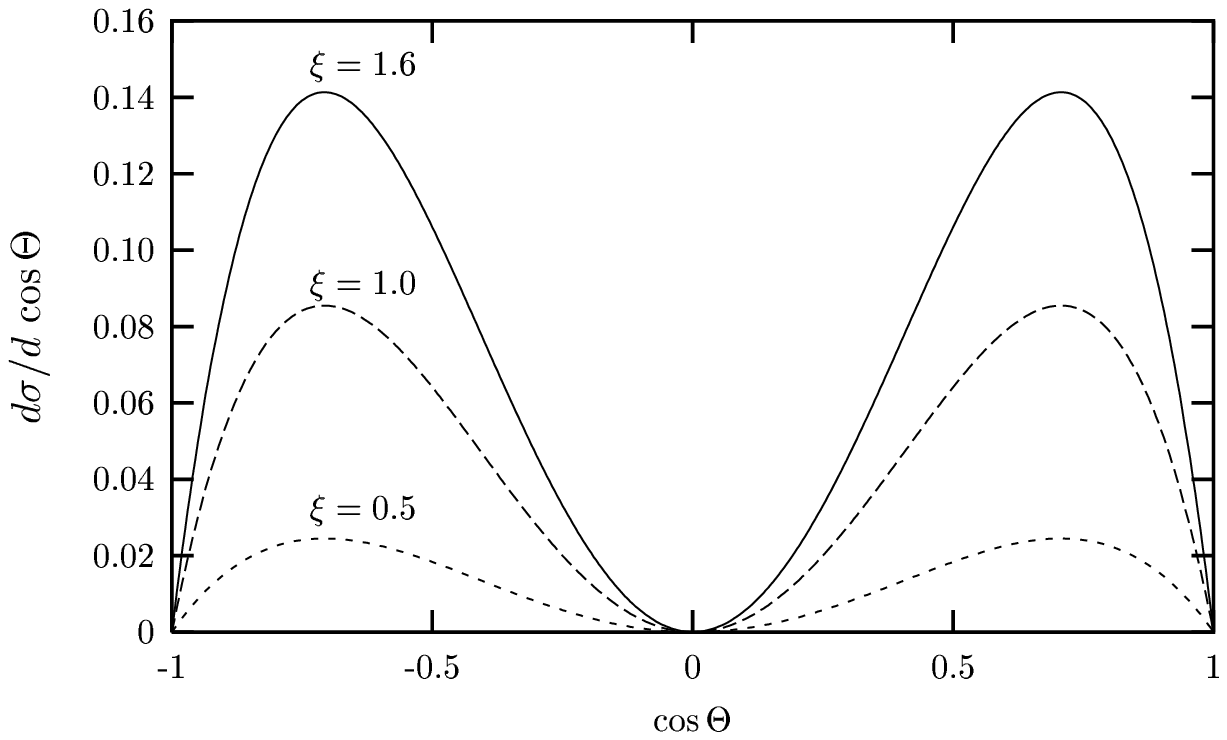}
\caption{\label{figdsdz} The $d \sg /d \cos\Theta$ in fb as a function of
$\cos\Theta$ for $\xi=0.5$, 1, and 1.6.
We set $\mphi=70$ GeV with $\Lam_\phi=5$ TeV and $m_0/\mpl=0.1$.}
\end{figure}

The differential cross section is
\beq
\label{diffsig}
\frac{d \sg}{d \cos\Theta}
=
\frac{g_{_{G h \phi}}^2\beta^{5/2}}{256 \pi s}
\left(
\frac{s}{\lwh^2}
\right)^2
\left(\sum_n \frac{1}{1 -{ m_G^{(n)\,2}}/s}
\right)^2 \sin^22\Theta
\,.
\eeq
In the above equation, the sum is over the KK states of the graviton.
With the input parameters, the mass $m_G^{(1)}$ of the first KK state is
about 782 GeV and $m_G^{(2)}$ is about 1.43 TeV.
Although the sum is over all states, the majority of the contributions
comes from the first state.
Figure \ref{figdsdz} presents
$d \sg /d \cos\Theta$ in fb as a function of
$\cos\Theta$ for $\xi=0.5$, 1, and 1.6.
We have set $\mphi=70$ GeV.
The angular distribution proportional to $\sin^2 2\Theta$
is characteristic of the spin-2 KK graviton exchange,
which is useful to
discriminate from any new vector-boson-exchange or scalar-exchange
contributions.

\begin{figure}
\includegraphics{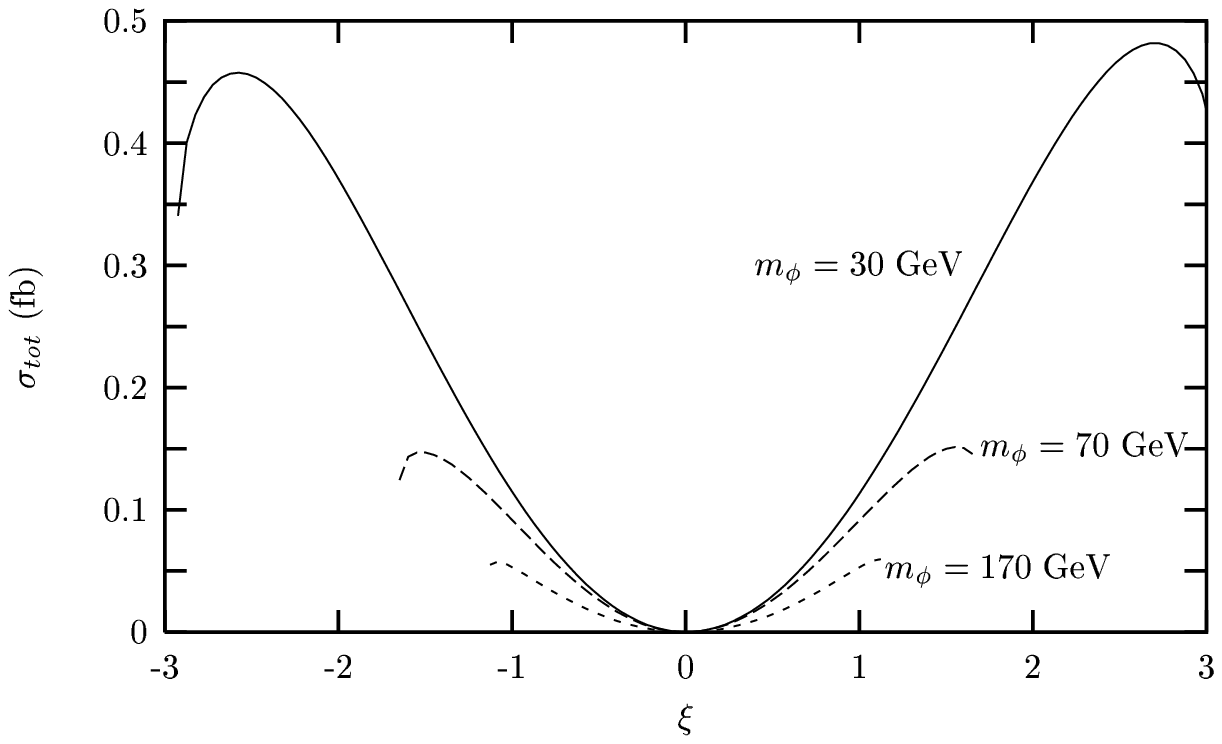}
\caption{\label{figtot} The total cross section $ \sg_{tot} $
in fb as a function of
$\xi$ for $m_\phi=30$, 70, and 170 GeV.}
\end{figure}

The total cross section is
\beq
\sg_{tot} =
\frac{g_{_{G h \phi}}^2\beta^{5/2}}{240 \pi s}
\left(
\frac{s}{\lwh^2}
\right)^2
\left(\sum_n \frac{1}{1 - { m_G^{(n)\,2}}/s}
\right)^2
\,.
\eeq
In Fig.~\ref{figtot},
we present the total cross section in fb as a function of
$\xi$ for $\mphi=30,~70,~170$ GeV
in the parameter space of $\xi$ allowed in
Eq.~(\ref{rootconstraint}).
The $\sig_{tot}$ decreases with increasing $\mphi$
due to kinematic suppression.
With the anticipated luminosity of 1000 fb$^{-1}$,
we have about a hundred events for $e^+ e^- \to h \phi$
if the mixing parameter $\xi$ is of order one.

Some discussions on the SM backgrounds are in order here.
Since the radion and Higgs boson decay promptly,
the final state will consist of at least four particles.
For $m_h=120$ GeV,
the radion-Higgs mixing does not change the dominant decay mode of the Higgs
boson of $h\to b\bar{b}$\,\cite{Gunion,Datta-HR-LHC}.
For the radion with mass $m_\phi \lsim 2 m_W$,
the decay mode into $ b\bar{b}$ is dominant or second dominant,
depending on the value of $\xi$.
Therefore, the final state will likely be $b\bar b b \bar b$.
We consider the SM background of
$e^+ e^- \to b \bar{b} b \bar{b}$.
In order to regulate the collinear divergences and afford realistic
detection criteria, we impose the selection cuts
  \beq
   p_{T_i} > 10 ~\mathrm{GeV},\quad
   | \cos\theta_i | < 0.95,\quad
     \cos(\theta_{ij}) < 0.9,
\eeq
where $p_{T_i}$ is the transverse momentum of each jet,
$\theta_i$ is the scattering angle of the final particle $i$,
and $\theta_{ij}$ is the angle between the particles $i$ and $j$.
Additional cut on the invariant mass of two jets
such as $ m_{h,\phi} - 5\,{\rm GeV} < m_{b\bar{b}}< m_{h,\phi} + 5\,{\rm GeV}$
will further reduce the background cross section
to the order of 0.001 fb at $\sqrt{s}=500$ GeV.
The SM background is almost negligible.
If the process $e^- e^+ \to h \phi$
has a total cross section of order of 0.01 fb,
the future linear colliders (LC) will be able to probe the process
exclusively sensitive to the radion-Higgs mixing.

\begin{figure}
\includegraphics{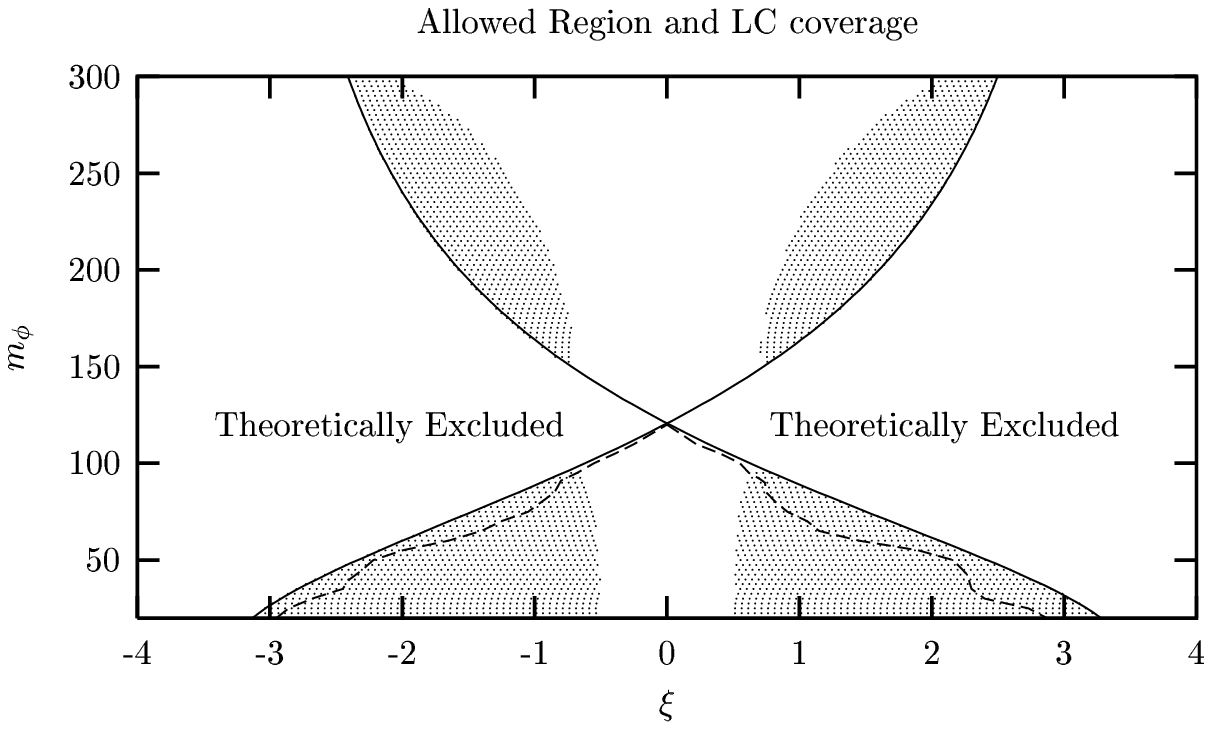}
\caption{\label{figconstraint} The allowed region of
$(\xi,\mphi)$ by the algebraic constraints
(inside the solid line)
and the LEP/LEP2 data
(inside the dashed line).
The Higgs mass $m_h$ is set to be 120 GeV.
For $m_\phi>m_h$,
the dashed line almost overlaps the solid line.
The dotted regions are for $\sig_{tot}(e^+ e^-\to h \phi)>0.03 $ fb.}
\end{figure}

Figure \ref{figconstraint} summarizes the allowed parameter space of
$(\xi,\mphi)$.
First,
the algebraic constraints in Eqs.~(\ref{rootconstraint}) and (\ref{xilim})
exclude
substantial regions denoted by `Theoretically excluded'
in Fig. \ref{figconstraint}.
The LEP and LEP II data exclude the region between the solid
and dashed line.
The dotted regions correspond to
$\sig_{tot}(e^+ e^-\to h \phi)>0.03 $ fb.
We have chosen a rather conservative criteria for sensitivity reach
for such a small SM background.
If $m_\phi<m_h$,
the process of $e^+ e^- \to \grav \to h \,\phi$ at the future LC
can probe
the parameter space with $|\xi| \gsim 0.5$.
Even when $m_\phi>m_h$, although the cross section is limited by the kinematics,
LC will be able to cover
a quite substantial portion of the parameter space
which the LEP/LEP II data cannot constrain.

If the Higgs boson mass is in fact heavier, say, $150$ GeV, the
sensitivity regions in the $(\xi,m_\phi)$ plane will be reduced,
especially for the region $m_\phi>m_h$.
An obvious reason is that the allowed kinematic phase space is reduced.
Whereas in the region $m_\phi<m_h$,
the sensitivity regions stay about the same.

\section{Conclusions}
\label{conclusion}

In the original Randall-Sundrum scenario in which all the SM
fields are confined on the visible brane,
the phenomenological signatures of the radion-Higgs mixing
at $e^+ e^-$ colliders
have been studied.
In the warped geometry,
the radion-Higgs mixing is weakly suppressed
by the VEV of the radion at electroweak scale.
It is known that the production and decay of the Higgs boson
are modified due to the radion-Higgs mixing
and thus the Higgs search strategy in the future colliders
needs refinement.
Complementarily,
high energy processes exclusively allowed for non-zero mixing
can also provide valuable information.
We pointed out that there are four triple-vertices which
would vanish without the radion-Higgs mixing.
In particular,
the vertex of $\grav$-$h$-$\phi$
has a large interaction strength, and involves
the KK graviton.
We studied the scattering process of
$e^+ e^- \to \grav \to h \phi$
at the future linear $e^+ e^-$ colliders.
The angular distribution
proportional to $\sin^2 2 \Theta$ has shown a unique feature of
the exchange of massive spin-2 gravitons.
Furthermore, in a large portion of the parameter space
where LEP/LEP II data cannot constrain,
the total cross section is large enough
for sensitivity reach.
We conclude that
$e^+ e^- \to \grav \to h \phi$
is a very sensitive process to probe the radion-Higgs mixing
in the Randall-Sundrum model.

\begin{acknowledgments}
The work of C.S.K. was supported by Grant No. 2001-042-D00022 of the KRF.
The work of J.S. was supported by Grant No. R02-2002-000-00168-0
from the Basic Research Program of the KOSEF. K.C. was supported by
NCTS under a grant from NSC, Taiwan.
\end{acknowledgments}

\end{document}